\documentclass[aps,prd,twocolumn,superscriptaddress]{revtex4}
\usepackage{epsfig,epsf}
\usepackage{amsmath}
\usepackage{amsthm}
\usepackage{amsfonts}
\usepackage{amssymb}
\usepackage{dsfont}
\usepackage{multirow}
\usepackage{appendix}
\usepackage{slashed}
\usepackage[active]{srcltx}
\usepackage{psfrag}

\begin{document}

\title{Decays of the light hybrid meson $1^{\mathrm{-+}}$ }
\date{\today}
\author{G.~Daylan Esmer}
\affiliation{Department of Physics, Istanbul University, Vezneciler, 34134 Istanbul, T%
\"{u}rkiye}
\author{K.~Azizi}
\thanks{Corresponding author: kazem.azizi@ut.ac.ir}
\affiliation{Department of Physics, University of Tehran, North Karegar Avenue, Tehran
14395-547, Iran}
\affiliation{Department of Physics, Do\v{g}u\c{s} University, Dudullu-\"{U}mraniye, 34775
Istanbul, T\"{u}rkiye}
\author{H.~Sundu}
\affiliation{Department of Physics Engineering, Istanbul Medeniyet University, 34700
Istanbul, T\"{u}rkiye}
\author{S.~T\"{u}rkmen}
\affiliation{Department of Physics, Istanbul University, Vezneciler, 34134 Istanbul, T%
\"{u}rkiye}

\begin{abstract}
The full width of the light isovector hybrid meson $H_{\mathrm{V}}$ with
spin-parities $1^{\mathrm{-+}}$ and content $(\overline{u}gu-\overline{d}gd)/%
\sqrt{2}$ is evaluated by considering the decays $H_{\mathrm{V}} \to
\rho^{\pm}\pi^{\mp}$, $b_1^{\pm}\pi^{\mp}$, $f_1(1285)\pi$, $f_1(1420)\pi$, $%
\eta \pi$, and $\eta^{\prime} \pi$. To calculate the partial widths of these
channels, we use QCD three-point sum rule method which is necessary to
determine strong couplings at the corresponding hybrid-meson-meson vertices.
It turns out that the main contribution to the full width $\Gamma[H_{\mathrm{%
V}}]=(109.7 \pm 16.0)~\mathrm{MeV}$ of the hybrid meson comes from the
processes $H_{\mathrm{V}} \to \rho^{\pm}\pi^{\mp}$ partial width of which
amounts to $\approx 67~\mathrm{MeV}$. The effects of the decays $H_{\mathrm{V%
}} \to b_1\pi $ and $H_{\mathrm{V}} \to f_1\pi, f_1^{\prime} \pi$ are also
sizeable: Their partial widths are equal to $13~\mathrm{MeV}$ and $20~%
\mathrm{MeV}$, respectively. The decays to $\eta \pi$ and $\eta^{\prime} \pi$
mesons are subdominant reactions, nevertheless they form $\approx 9\%$ of
the full width $\Gamma[H_{\mathrm{V}}]$. Results obtained in this work may
be interesting to unravel the tangle of predictions about $H_{\mathrm{V}}$
existing in the literature, as well as useful in analyses of different
resonances.
\end{abstract}

\maketitle


\section{Introduction}

\label{sec:intro}

During last decades, the hadron spectroscopy was replenished with new
particles and resonances discovered by experimental collaborations in
different processes. Some of them were interpreted as excited states of
mesons and baryons observed and identified already in the experiments.
Others may be considered as evidences for conventional hadrons seen for the
first time. But there are resonances which can not be included into the
standard picture of $q\overline{q}^{\prime }$ mesons and $qq^{\prime
}q^{\prime \prime }$ baryons. They may have diquark-antidiquark, pentaquark,
glueball or hybrid structures and belong to the class of unusual or exotic
hadrons.

Various hybrid mesons containing apart from valence quarks also a valence
gluon field are interesting objects for the experimental and theoretical
studies. Actually such particles attracted interest of researches almost
fifty years ago starting from the first days of the parton model and quantum
chromodynamics (QCD) \cite{Jaffe:1975fd}. Theoretical investigations of
hybrid states have rather long history and encompass numerous publications
and results obtained in the framework of various methods and models \cite%
{Meyer:2015eta} (and references therein).

It is worth to emphasize that the hybrid mesons can carry quantum numbers
which are forbidden for the conventional mesons. For instance, they can have
spin-parities $J^{\mathrm{PC}}={0^{--},~0^{+-},~1^{-+},~2^{-+}}$ which are
not allowed for the mesons with quark-antiquark structure. Therefore,
discovery of the particles with such quantum numbers provides valuable
knowledge about the exotic mesons and features of the low energy QCD.

Information on structures which can be considered as the potential
candidates to hybrid mesons became available since the experiments reported
in Ref.\ \cite{IHEP-Brussels-LosAlamos-AnnecyLAPP:1988iqi}, where ${1^{-+}}$
meson with the mass $(1406\pm 20)~\mathrm{MeV}$ and width $(180\pm 30)~%
\mathrm{MeV}$ was observed in the exclusive reaction $\pi ^{-}p\rightarrow
\pi ^{0}n\eta $. It was labeled as $\pi _{1}(1400)$ and became the first
member of the ${1^{-+}}$ mesons' family. The structure ${1^{-+}}$ was
explored in other experiments as well \cite%
{CrystalBarrel:1998cfz,E852:2004gpn,E852:2004rfa,COMPASS:2018uzl,COMPASS:2021ogp}%
. Data collected and critically analyzed during this process led to the
picture: There are two isovector resonances $\pi _{1}(1400)$ and $\pi
_{1}(1600)$ with $J^{\mathrm{PC}}={1^{-+}}$. But the coupled channel
analyses carried out in Refs.\ \cite{JPAC:2018zyd,Kopf:2020yoa} favor the
existence of only one broad state $\pi _{1}(1600)$ (see, also Ref.\ \cite%
{PDG:2024}).

The evidence for the next exotic meson $\pi _{1}(2015)$ from this family was
reported in Refs.\ \cite{E852:2004gpn,E852:2004rfa}. The isoscalar vector
particle $\eta _{1}(1855)$ with the quantum numbers ${1^{-+}}$ was seen
quite recently by the BESIII collaboration in the radiative decay $J/\psi
\rightarrow \gamma \eta \eta ^{\prime }$ \cite{BESIII:2022riz}.

The hybrid mesons of different contents and quantum numbers were explored in
the context of numerous theoretical methods. Thus, they were considered
using QCD sum rule (SR) method in the articles \cite%
{Balitsky:1982ps,Balitsky:1986hf,Latorre:1984kc,Govaerts:1985fx,Chetyrkin:2000tj,Narison:2009vj,Chen:2010ic, Huang:2010dc,Zhang:2013rya,Huang:2014hya,Huang:2016upt,Barsbay:2022gtu,Alaakol:2024zyh,Barsbay:2024vjt}.  As a candidate to a lightest hybrid meson, the state $\overline{q}gq$ with
exotic spin-parities $J^{\mathrm{PC}}={1^{-+}}$ was studied in a more
detailed form. But available predictions for the mass of this particle are
controversial and differ from each other considerably. Thus, in Ref.\ \cite%
{Balitsky:1986hf} the masses of the nonstrange and strange light hybrid
states ${1^{-+}}$ were found around of $1.5~\mathrm{GeV}$ and $1.6~\mathrm{%
GeV}$, respectively. The result $(1.7\pm 0.1)~\mathrm{GeV}$ was predicted in
Ref.\ \cite{Latorre:1984kc}. The comprehensive studies were carried out in
the sum rules framework for the light hybrid mesons with $J=0$, $1$ and
different combinations of the quantum numbers $\mathrm{P}$ and $\mathrm{C}$
as well \cite{Govaerts:1985fx}. The mass of the hybrid ${1^{-+}}$ was found
equal to $(1.6-1.7)~\mathrm{GeV}$, $1.81(6)~\mathrm{GeV}$, $(1.71\pm 0.22)~%
\mathrm{GeV}$ and $(1.72-2.60)~\mathrm{GeV}$ in Refs.\ \cite%
{Chetyrkin:2000tj,Narison:2009vj,Zhang:2013rya,Huang:2014hya}, respectively.
Recently, the SR studies of the light hybrid meson $1^{-+}$ were updated by
including into analysis quark-gluon condensates up to ten dimensions: The
mass of this particle amounts to $2.30_{-0.17}^{+0.18}~\mathrm{GeV}$ \cite%
{Barsbay:2024vjt}.

The parameters of the light hybrids, their production mechanisms and decay
channels were explored using alternative theoretical approaches as well \cite%
{Lacock:1996ny,McNeile:2006bz,Burns:2006wz,Burden:2002ps,Close:1994hc,Barnes:1995hc,Barnes:1982zs,Chanowitz:1982qj,Bellantuono:2014lra,Swanson:2023zim,Farina:2023oqk}%
. In fact, lattice computations predicted the existence of the hybrid meson $%
{1^{-+}}$ with the mass $(1.9\pm 0.2)~\mathrm{GeV}$ \cite{Lacock:1996ny}.
The two exotic states with masses $1439~\mathrm{MeV}$ and $1498~\mathrm{MeV}$
were found based on the Dyson-Schwinger equations in Ref.\ \cite%
{Burden:2002ps}. Relevant investigations were also carried out in the
context of the flux tube \cite{Close:1994hc,Barnes:1995hc}, the MIT bag \cite%
{Barnes:1982zs,Chanowitz:1982qj}, a holographic QCD \cite%
{Bellantuono:2014lra} and the constituent gluon models \cite%
{Swanson:2023zim,Farina:2023oqk}.

The decays of the hybrid state $J^{\mathrm{PC}}={1^{-+}}$ and its full width
is also the important problem which was addressed in the context of various
methods. It is known that the meson $\pi _{1}(1600)$ is the broad particle
with the width $370_{-60}^{+50}~\mathrm{MeV}$ \cite{PDG:2024}. Having
suggested a hybrid interpretation for $\pi _{1}(1600)$ one should explain
its decay modes as well. Theoretical analyses of this question led to
interesting and sometimes to mutually exclusive conclusions. For instance,
the lattice simulations and flux tube model predicted a prevalence of the
decay to $b_{1}\pi $ mesons as a main channel for the transformation of the
hybrid state ${1^{-+}}$ \cite{McNeile:2006bz,Burns:2006wz}, whereas in the
SR computations the process ${1^{-+}\rightarrow }\rho \pi $ was found to be
the dominant one \cite{Chen:2010ic}. There are also other decay channels of
the meson $\pi _{1}(1600)$ reported in Ref.\ \cite{PDG:2024} and considered
in the literature.

In the current work, we explore decays of the isovector hybrid meson $J^{%
\mathrm{PC}}={1^{-+}}$ by applying the QCD sum rule approach \cite%
{Shifman:1978bx, Shifman:1978by}. This method is one of the powerful and
predictive nonperturbative tools for calculations of hadrons' parameters.
Originally elaborated to explore conventional mesons and baryons, it was
successfully employed to study also the exotic hadrons \cite%
{Albuquerque:2018jkn,Agaev:2020zad}.

In our analysis we rely on results for the mass and current coupling of this
state obtained in Ref.\ \cite{Barsbay:2024vjt} and, in what follows, denote
it as $H_{\mathrm{V}}$. We are going to consider the decays $H_{\mathrm{V}%
}\rightarrow \rho \pi $, $b_{1}\pi $, $f_{1}(1285)\pi $, $f_{1}(1420)\pi $, $%
\eta \pi $, and $\eta ^{\prime }\pi $ of this particle. For simplicity of
the presentation, we will employ notations $f_{1}$ and $f_{1}^{\prime }$ for
the mesons $f_{1}(1285)$ and $f_{1}(1420)$, respectively. Partial widths of
these processes depend on the strong couplings $g_{i}$ at the
hybrid-meson-meson vertices. We determine them by employing the three-point
sum rule method which allow us to find the sum rules for the corresponding
form factors $g_{i}(q^{2})$. At the mass shell of a final pion $q^{2}=m_{\pi
}^{2}$ these form factors give the required strong couplings.

We divide the article into the following sections: In Sec.\ \ref{sec:Decays1}%
, we consider the processes $H_{\mathrm{V}}\rightarrow $ $\rho ^{-}\pi ^{+}$
and $H_{\mathrm{V}}\rightarrow \rho ^{+}\pi ^{-}$. We evaluate the coupling
at the vertex $H_{\mathrm{V}}\rho ^{-}\pi ^{+}$ and find the partial width
of this decay. The partial width of the second process $H_{\mathrm{V}%
}\rightarrow \rho ^{+}\pi ^{-}$ is, evidently, equal to the width of the
first one. In Sec.\ \ref{sec:Decays2}, we concentrate on the channels $H_{%
\mathrm{V}}\rightarrow $ $b_{1}\pi $, $f_{1}\pi $, and $f_{1}^{\prime }\pi $
and calculate their partial widths. The decays to $\eta \pi $ and $\eta
^{\prime }\pi $ mesons are studied in Sec.\ \ref{sec:Decays3}. Here, we also
find the full width of the hybrid meson $H_{\mathrm{V}}$. In the last part
of the article, we compare our predictions with results of various works
obtained by means of SR or other methods.


\section{Decays $H_{\mathrm{V}}\rightarrow \protect\rho \protect\pi $}

\label{sec:Decays1}

In this section, we explore the decays $H_{\mathrm{V}}\rightarrow \rho
^{-}\pi ^{+}$ and $H_{\mathrm{V}}\rightarrow \rho ^{+}\pi ^{-}$ of the
hybrid isovector meson $H_{\mathrm{V}}$. We are going to present in a
detailed form calculation of the process $H_{\mathrm{V}}(p)\rightarrow \rho
^{-}(p^{\prime })\pi ^{+}(q)$, where $p$, $p^{\prime }$ and $q$ are momenta
of the particles involved into the decay.

The coupling $g_{1}$ which describes the strong interaction of particles at
the vertex $H_{\mathrm{V}}\rho ^{-}\pi ^{+}$ is equal to the form factor $%
g_{1}(q^{2})$ at the mass shell of the pion $q^{2}=m_{\pi }^{2}$. In its
turn, the sum rule for $g_{1}(q^{2})$ can be obtained from analysis of the
three-point correlation function.

To this end, we consider the three-point correlator given by the expression
\begin{eqnarray}
\Pi _{\mu \nu }(p,p^{\prime }) &=&i^{2}\int d^{4}xd^{4}ye^{ip^{\prime
}x+iyq}\langle 0|\mathcal{T}\{J_{\nu }^{\rho }(x)  \notag \\
&&\times J^{\pi }(y)J_{\mu }^{\dagger }(0)\}|0\rangle ,  \label{eq:CorrF1}
\end{eqnarray}%
where $J_{\mu }(x)$ is the interpolating current for the vector hybrid
state,
\begin{eqnarray}
J_{\mu }(x) &=&\frac{1}{\sqrt{2}}g_{s}\frac{\lambda _{ab}^{n}}{2}G_{\mu
\theta }^{n}(x)\left[ \overline{u}_{a}(x)\gamma ^{\theta }u_{b}(x)\right.
\notag \\
&&\left. -\overline{d}_{a}(x)\gamma ^{\theta }d_{b}(x)\right] .
\label{eq:CR1}
\end{eqnarray}%
Here, $g_{s}$ is the strong coupling constant, $a,b=1,2,3$ and $n=1,2,\cdots
,8$ are color indices, $\lambda ^{n}$ and $G_{\mu \theta }^{n}(x)$ are the
Gell-Mann matrices and the gluon field strength tensor, respectively.

The currents $J_{\nu }^{\rho }(x)$ and $J^{\pi }(x)$ for the conventional
mesons $\rho ^{-}$ and $\pi ^{+}$ have the following forms%
\begin{equation}
J_{\nu }^{\rho }(x)=\overline{u}_{i}(x)\gamma _{\nu }d_{i}(x),\ J^{\pi }(x)=%
\overline{d}_{j}(x)i\gamma _{5}u_{j}(x),
\end{equation}%
with $i,j=1,2,3$ being the color indices.

To derive the sum rule for the form factor $g_{1}(q^{2})$ one needs to write
the correlator $\Pi _{\mu \nu }(p,p^{\prime })$ by employing the
phenomenological parameters of the particles $H_{\mathrm{V}}$, $\rho ^{-}$
and $\pi ^{+}$, and also express it using the quark and gluon propagators.
The first expression $\Pi _{\mu \nu }^{\mathrm{Phys}}(p,p^{\prime })$
establishes the physical side of the sum rule equality, whereas the second
one $\Pi _{\mu \nu }^{\mathrm{OPE}}(p,p^{\prime })$ is the QCD side of this
equality.

The $\Pi _{\mu \nu }^{\mathrm{Phys}}(p,p^{\prime })$ has the following form%
\begin{eqnarray}
&&\Pi _{\mu \nu }^{\mathrm{Phys}}(p,p^{\prime })=\frac{\langle 0|J_{\nu
}^{\rho }|\rho ^{-}(p^{\prime },\varepsilon ^{\prime })\rangle \langle
0|J^{\pi }|\pi ^{+}(q)\rangle }{(p^{\prime 2}-m_{\rho }^{2})(q^{2}-m_{\pi
}^{2})}  \notag \\
&&\times \frac{\langle \rho ^{-}(p^{\prime },\varepsilon ^{\prime })\pi
^{+}(q)|H_{\mathrm{V}}(p,\varepsilon )\rangle \langle H_{\mathrm{V}%
}(p,\varepsilon )|J_{\mu }^{\dagger }|0\rangle }{(p^{2}-m^{2})}+\cdots , \notag \\
\label{eq:CorrF2}
\end{eqnarray}%
where $m$,\ $m_{\rho }$ and $m_{\pi }$ are the masses of the hybrid state $%
H_{\mathrm{V}}$ and mesons $\rho ^{-}$ and $\pi ^{+}$, respectively. Here, $%
\varepsilon _{\mu }$ and $\varepsilon _{\nu }^{\prime }$ are the
polarization vectors of the particles $H_{\mathrm{V}}$ and $\rho ^{-}$. The
term written down explicitly is the contribution of the ground-level
particles: Effects of higher excited states and continuum are shown by the
dots.

The correlation function $\Pi _{\mu \nu }^{\mathrm{Phys}}(p,p^{\prime })$
can be rewritten in a more convenient form. For these purposes, we introduce
the matrix elements
\begin{eqnarray}
\langle 0|J_{\nu }^{\rho }|\rho ^{-}(p^{\prime },\varepsilon ^{\prime
})\rangle  &=&m_{\rho }f_{\rho }\varepsilon _{\nu }^{\prime },\ \langle
0|J^{\pi }|\pi ^{+}(q)\rangle =\frac{f_{\pi }m_{\pi }^{2}}{2m_{q}},  \notag
\\
\langle 0|J_{\mu }|H_{\mathrm{V}}((p,\varepsilon ))\rangle  &=&mf\varepsilon
_{\mu },  \label{eq:ME1}
\end{eqnarray}%
and
\begin{equation}
\langle \rho ^{-}(p^{\prime },\varepsilon ^{\prime })\pi ^{+}(q)|H_{\mathrm{V%
}}(p,\varepsilon )\rangle =g_{1}(q^{2})\epsilon _{\sigma \tau \alpha \beta
}q^{\alpha }p^{\prime \beta }\varepsilon ^{\sigma }\varepsilon ^{\prime \ast
\tau }.  \label{eq:ME2}
\end{equation}%
In Eq.\ (\ref{eq:ME1}), $2m_{q}=m_{u}+m_{d}$, whereas $f$,\ $f_{\rho }$ and $%
f_{\pi }$ are the current coupling and decay constants of the hybrid $H_{%
\mathrm{V}}$ and the mesons $\rho ^{-}$ and $\pi ^{+}$, respectively. The
matrix element $\langle 0|J^{\pi }|\pi ^{+}(q)\rangle $ can also be written
using the parameter $\mu _{\pi }$
\begin{equation}
\mu _{\pi }=\frac{m_{\pi }^{2}}{2m_{q}}=-\frac{\langle \overline{q}q\rangle
}{f_{\pi }^{2}}.  \label{eq:Eq1}
\end{equation}%
The second equality in Eq.\ (\ref{eq:Eq1}) is the relation between $m_{\pi }$%
, $f_{\pi }$, quark masses and the quark condensate $\langle \overline{q}%
q\rangle $ arising from the partial conservation of the axial vector current
(PCAC).

Then the correlator $\Pi _{\mu \nu }^{\mathrm{Phys}}(p,p^{\prime })$ amounts
to
\begin{eqnarray}
\Pi _{\mu \nu }^{\mathrm{Phys}}(p,p^{\prime }) &=&\frac{g_{1}(q^{2})mfm_{%
\rho }f_{\rho }f_{\pi }m_{\pi }^{2}}{2m_{q}(p^{\prime 2}-m_{\rho
}^{2})(p^{2}-m^{2})}  \notag \\
&&\times \frac{\epsilon _{\alpha \beta \mu \nu }q^{\alpha }p^{\prime \beta }%
}{(q^{2}-m_{\pi }^{2})}\cdots .  \label{eq:PhysS1}
\end{eqnarray}%
The function $\Pi _{\mu \nu }^{\mathrm{Phys}}(p,p^{\prime })$ contains only
one term proportional to the structure $\epsilon _{\alpha \beta \mu \nu
}q^{\alpha }p^{\prime \beta }$. The corresponding function
\begin{eqnarray}
\Pi ^{\mathrm{Phys}}\left( p^{2},p^{\prime 2},q^{2}\right)  &=&\frac{%
g_{1}(q^{2})mfm_{\rho }f_{\rho }}{2m_{q}(p^{\prime 2}-m_{\rho
}^{2})(p^{2}-m^{2})}  \notag \\
&&\times \frac{f_{\pi }m_{\pi }^{2}}{(q^{2}-m_{\pi }^{2})},
\label{eq:PhysS1A}
\end{eqnarray}%
is the invariant amplitude which will be employed in the following
consideration.

It can be presented through double dispersion integral \cite%
{Ioffe:1982qb,Colangelo:2000dp}%
\begin{eqnarray}
&&\Pi ^{\mathrm{Phys}}(p^{2},p^{\prime 2},q^{2})=\frac{\widehat{g}_{1}(q^{2})%
}{\left( p^{2}-m^{2}\right) (p^{\prime 2}-m_{\rho }^{2})}  \notag \\
&&+\int \int_{\Sigma }dsds^{\prime }\frac{\rho ^{\mathrm{h}}(s,s^{\prime
},q^{2})}{(s-p^{2})(s^{\prime }-p^{\prime 2})}+\cdots .  \label{eq:PhysS2}
\end{eqnarray}%
Here, $\widehat{g}_{1}(q^{2})$ contains input information from Eq.\ (\ref%
{eq:PhysS1}), whereas the second term is contributions of the excited and
continuum states. The function $\rho ^{\mathrm{h}}(s,s^{\prime },q^{2})$ is
unknown hadronic double spectral density. The $\Sigma $ is a region in the $%
(s,s^{\prime })$ plane and has the boundaries $(s_{0},s_{0}^{\prime })$
which depend on a process under consideration. The single dispersion
integrals over $s$ and $s^{\prime }$ which vanish after the double Borel
transformations are shown in Eq.\ (\ref{eq:PhysS2}) by the ellipses. The
Eq.\ (\ref{eq:PhysS2}) can be obtained from the dispersion integral%
\begin{equation}
\Pi ^{\mathrm{Phys}}(p^{2},p^{\prime 2},q^{2})=\int \int_{\Sigma
}dsds^{\prime }\frac{\rho ^{\mathrm{Phys}}(s,s^{\prime },q^{2})}{%
(s-p^{2})(s^{\prime }-p^{\prime 2})},
\end{equation}%
using the double spectral density
\begin{eqnarray}
&&\rho ^{\mathrm{Phys}}(s,s^{\prime },q^{2})=\widehat{g}_{1}(q^{2})\delta
(s-m^{2})\delta (s^{\prime }-m_{\rho }^{2})  \notag \\
&&+\rho ^{\mathrm{h}}(s,s^{\prime },q^{2})\theta (s-s_{0})\theta (s^{\prime
}-s_{0}^{\prime }),
\end{eqnarray}%
where,  $\theta (z)$ is the unit step function.

To find the SR for the form factor $g_{1}(q^{2})$, the same correlation
function $\Pi _{\mu \nu }(p,p^{\prime })$ has to be expressed in terms of
the light quark propagators and computed with some accuracy in the operator
product expansion ($\mathrm{OPE}$). For $\Pi _{\mu \nu }^{\mathrm{OPE}%
}(p,p^{\prime })$ calculations lead to the result%
\begin{eqnarray}
&&\Pi _{\mu \nu }^{\mathrm{OPE}}(p,p^{\prime })=\frac{g_{s}\lambda _{ab}^{n}%
}{2\sqrt{2}}i\int d^{4}xd^{4}ye^{ip^{\prime }x+iyq}G_{\mu \theta }^{n}(0)
\notag \\
&&\times \left\{ \mathrm{Tr}\left[ \gamma _{\nu }S_{d}^{ij}(x-y)\gamma
_{5}S_{u}^{jb}(y)\gamma ^{\theta }S_{u}^{ai}(-x)\right] \right.  \notag \\
&&\left. -\mathrm{Tr}\left[ \gamma _{\nu }S_{d}^{ib}(x)\gamma ^{\theta
}S_{d}^{aj}(-y)\gamma _{5}S_{u}^{ji}(y-x)\right] \right\} .  \label{eq:OPE1}
\end{eqnarray}%
In Eq.\ (\ref{eq:OPE1}) $S_{d(u)}^{ab}(x)$ are the propagators of the $u$
and $d$ quarks%
\begin{eqnarray}
&&S_{u(d)}^{ab}(x)=i\delta _{ab}\frac{\slashed x}{2\pi ^{2}x^{4}}-\delta
_{ab}\frac{m_{q}}{4\pi ^{2}x^{2}}-\delta _{ab}\frac{\langle \overline{q}%
q\rangle }{12}  \notag \\
&&+i\delta _{ab}m_{u(d)}\frac{\slashed x\langle \overline{q}q\rangle }{48}%
-\delta _{ab}\frac{x^{2}}{192}\langle \overline{q}g_{s}\sigma Gq\rangle
\notag \\
&&+i\delta _{ab}m_{u(d)}\frac{x^{2}\slashed x}{1152}\langle \overline{q}%
g_{s}\sigma Gq\rangle -i\frac{g_{s}G_{ab}^{\alpha \beta }}{32\pi ^{2}x^{2}}%
\left[ \slashed x{\sigma _{\alpha \beta }+\sigma _{\alpha \beta }}\slashed x%
\right]  \notag \\
&&-i\delta _{ab}\frac{x^{2}\slashed xg_{s}^{2}\langle \overline{q}q\rangle
^{2}}{7776}-\delta _{ab}\frac{x^{4}\langle \overline{q}q\rangle \langle
g_{s}^{2}G^{2}\rangle }{27648}+\cdots ,  \label{eq:Prop}
\end{eqnarray}%
where $G_{\alpha \beta }^{ab}(x)$ is
\begin{equation}
G_{\alpha \beta }^{ab}(x)=G_{\alpha \beta }^{m}(x)\lambda _{ab}^{m}/2.
\end{equation}%
In our calculations we adopt for Eq.\ (\ref{eq:Prop}) the approximation $%
m_{u}=m_{d}=0$. The correlation function $\Pi _{\mu \nu }^{\mathrm{OPE}%
}(p,p^{\prime })$ contains the vacuum condensates generated by the terms in
Eq.\ (\ref{eq:Prop}) which are not the full contributions.

With expressions Eqs.\ (\ref{eq:OPE1}) and (\ref{eq:Prop}) at hand, we can
explain computation of the QCD side of the sum rule. As is seen, $\Pi _{\mu
\nu }^{\mathrm{OPE}}(p,p^{\prime })$ contains three light quark propagators
and the tensor $G_{\mu \theta }^{n}(0)$. This gluon field strength tensor
contracted with the term $-ig_{s}G_{\alpha \beta }^{ab}(x)\left[ \slashed %
x\sigma ^{\alpha \beta }+\sigma ^{\alpha \beta }\slashed x\right] /32\pi
^{2}x^{2}$ from the quark propagators gives rise to the matrix element of
two-gluon fields sandwiched between the vacuum states $\langle 0|G_{\alpha
\beta }^{m}(x)G_{\mu \theta }^{n}(0)|0\rangle $. This two-gluon matrix
element is treated by two different manners. Thus, first we consider it as
the full gluon propagator in coordinate space between two points $0$ and $x$%
, and use the formula
\begin{eqnarray}
&&\langle 0|G_{\alpha \beta }^{m}(x)G_{\mu \theta }^{n}(0)|0\rangle =\frac{%
\delta ^{mn}}{2\pi ^{2}x^{4}}\left[ g_{\beta \theta }\left( g_{\alpha \mu }-%
\frac{4x_{\alpha }x_{\mu }}{x^{2}}\right) \right.  \notag \\
&&+(\beta ,\theta )\leftrightarrow (\alpha ,\mu )-\beta \leftrightarrow
\alpha -\theta \leftrightarrow \mu ].
\end{eqnarray}%
The matrix element $\langle 0|G_{\alpha \beta }^{m}(x)G_{\mu \theta
}^{n}(0)|0\rangle $ is also treated as the two-gluon condensate. To this
end, the gluon field at point $x$ is expanded around $x=0$ and the first
term is taken into account. As a result, we get
\begin{eqnarray}
&&\langle 0|g_{s}^{2}G_{\alpha \beta }^{m}(x)G_{\mu \theta }^{n}(0)|0\rangle
=\frac{\langle g_{s}^{2}G^{2}\rangle }{96}\delta ^{mn}[g_{\alpha \mu
}g_{\beta \theta }  \notag  \label{eq:Gcond} \\
&&-g_{\alpha \theta }g_{\mu \beta }].
\end{eqnarray}%
After these manipulations, one multiplies these terms to remaining two light
quark propagators and other factors and finishes calculations.

The correlation function $\Pi _{\mu \nu }^{\mathrm{OPE}}(p,p^{\prime })$ has
the same Lorentz structure as $\Pi _{\mu \nu }^{\mathrm{Phys}}(p,p^{\prime
}) $. Having denoted the corresponding invariant amplitude by $\Pi ^{\mathrm{%
OPE}}(p^{2},p^{\prime 2},q^{2})$ and applied the double dispersion relation,
we get
\begin{equation}
\Pi ^{\mathrm{OPE}}(p^{2},p^{\prime 2},q^{2})=\int_{0}^{\infty
}\int_{0}^{\infty }\frac{dsds^{\prime }\rho ^{\mathrm{OPE}}(s,s^{\prime
},q^{2})}{(s-p^{2})(s^{\prime }-p^{\prime 2})}+\cdots .  \label{eq:DDR}
\end{equation}%
Note that the spectral density $\rho ^{\mathrm{OPE}}(s,s^{\prime },q^{2})$
is equal to a double discontinuity (an imaginary part) of $\Pi ^{\mathrm{OPE}%
}(s,s^{\prime },q^{2})$.

After equating the functions $\Pi ^{\mathrm{Phys}}(p^{2},p^{\prime 2},q^{2})$
and $\Pi ^{\mathrm{OPE}}(p^{2},p^{\prime 2},q^{2})$, performing the double
Borel transformations over the variables $-p^{2}$, $-p^{\prime 2}$ and
subtracting from the QCD side of this equality contributions of the excited
and continuum states under the quark-hadron duality assumption $\rho ^{%
\mathrm{h}}(s,s^{\prime },q^{2})\simeq \rho ^{\mathrm{OPE}}(s,s^{\prime
},q^{2})$, we find the sum rule for $g_{1}(q^{2})$
\begin{eqnarray}
&&g_{1}(q^{2})=\frac{2m_{q}(q^{2}-m_{\pi }^{2})}{mfm_{\rho }f_{\rho }f_{\pi
}m_{\pi }^{2}}e^{m^{2}/M_{1}^{2}}e^{m_{\rho }^{2}/M_{2}^{2}}  \notag \\
&&\times \Pi (\mathbf{M}^{2},\mathbf{s}_{0},q^{2}).  \label{eq:SRG}
\end{eqnarray}%
Here, the function $\Pi (\mathbf{M}^{2},\mathbf{s}_{0},q^{2})$ \ has the
form
\begin{eqnarray}
&&\Pi (\mathbf{M}^{2},\mathbf{s}_{0},q^{2})=\int_{0}^{s_{0}}ds%
\int_{0}^{s_{0}^{\prime }}ds^{\prime }\rho ^{\mathrm{OPE}}(s,s^{\prime
},q^{2})  \notag \\
&&\times e^{-s/M_{1}^{2}}e^{-s^{\prime }/M_{2}^{2}}+\Pi (\mathbf{M}^{2}).
\end{eqnarray}%
It contains the parameters $\mathbf{M}^{2}=(M_{1}^{2},M_{2}^{2})$ and $%
\mathbf{s}_{0}=(s_{0},s_{0}^{\prime })$ where the pairs $(M_{1}^{2},s_{0})$
and $(M_{2}^{2},s_{0}^{\prime })$ correspond to $H_{\mathrm{V}}$ and $\rho $
channels, respectively.

The spectral density $\rho ^{\mathrm{OPE}}(s,s^{\prime },q^{2})$ for the
decay $H_{\mathrm{V}}\rightarrow \rho \pi $ is given by the formula
\begin{eqnarray}
&&\rho ^{\mathrm{OPE}}(s,s^{\prime },q^{2})=\frac{g_{s}^{2}m_{q}}{32\sqrt{2}%
\pi ^{4}}\int_{0}^{1}d\alpha \int_{0}^{1-\alpha }\frac{d\beta }{\beta \alpha
(1-\alpha -\beta )}  \notag \\
&&\times \left[ \beta ^{2}+\alpha ^{2}(\alpha -1)+\beta (\alpha ^{2}+2\alpha
-1)\right] \theta (N)
\end{eqnarray}%
where $\alpha $ and $\beta $ are the Feynman parameters, and the argument $N$
of the function $\theta (N)$ amounts to
\begin{equation}
N=(s\beta +s^{\prime }\alpha )(1-\alpha -\beta )+q^{2}\beta \alpha .
\end{equation}
The nonperturbative term $\Pi (\mathbf{M}^{2})$ is determined directly from $%
\Pi ^{\mathrm{OPE}}(p^{2},p^{\prime 2},q^{2})$ through double Borel
transformations and has the form%
\begin{eqnarray}
&&\Pi (\mathbf{M}^{2})=-\frac{\pi ^{2}\langle \alpha _{s}G^{2}/\pi \rangle
\langle \overline{q}q\rangle }{18\sqrt{2}}  \notag \\
&&\times \int_{0}^{1}d\alpha \frac{1}{\alpha ^{2}M_{1}^{2}}\delta \left(
\frac{1}{M_{2}^{2}}-\frac{1-\alpha }{\alpha M_{1}^{2}}\right) .
\end{eqnarray}

The SR given in Eq.\ (\ref{eq:SRG}) depend on the masses and decay constants
of the particles involved into this decay. The parameters of the $\rho $ and
$\pi $ mesons are well known quantities \cite{PDG:2024}: We use $m_{\rho
}=(775.11\pm 0.34)~\mathrm{MeV,}$ $f_{\rho }=(216\pm 3)~\mathrm{MeV}$ and $%
m_{\pi }=(139.57039\pm 0.00017)~\mathrm{MeV,}$ $f_{\pi }=(130.2\pm 0.8)~%
\mathrm{MeV}$, and $m_{q}=(3.49\pm 0.07)~\mathrm{MeV}$. The mass $m=$ $%
2.30_{-0.17}^{+0.18}~\mathrm{GeV}$ and current coupling $f=\sqrt{2}\cdot
0.38_{-0.04}^{+0.05}\cdot 10^{-1}~\mathrm{GeV}^{3}$ of the vector hybrid $H_{%
\mathrm{V}}$ were evaluated in the sum rule framework in Ref.\ \cite%
{Barsbay:2024vjt}. Besides, the correlation function $\Pi (\mathbf{M}^{2},%
\mathbf{s}_{0},q^{2})$ depends on the condensates
\begin{eqnarray}
\langle \alpha _{s}G^{2}/\pi \rangle &=&(0.012\pm 0.004)~\mathrm{GeV}^{4},
\notag \\
\langle \overline{q}q\rangle &=&-(0.24\pm 0.01)^{3}~\mathrm{GeV}^{3},
\label{eq:Cond}
\end{eqnarray}%
which were extracted from analysis of the different hadronic processes \cite%
{Shifman:1978bx, Shifman:1978by}.

For numerical calculations, one also needs to fix parameters $\mathbf{M}^{2}$%
, $\mathbf{s}_{0}$. They should meet standard requirements of the sum rule
computations. Prevalence of the pole contribution to extracted quantities,
as well as relative stability of numerical predictions under variations of $%
\mathbf{M}^{2}=(M_{1}^{2},M_{2}^{2})$ are among such important constrains.
In the $H_{\mathrm{V}}$ channel, we choose \cite{Barsbay:2024vjt}
\begin{equation}
M_{1}^{2}\in \lbrack 2.5,3.5]~\mathrm{GeV}^{2},\ s_{0}\in \lbrack 8,10]~%
\mathrm{GeV}^{2}.  \label{eq:Wind1}
\end{equation}%
These intervals coincide with corresponding working windows used in the
calculation of the mass $m$ and current coupling $f$ of the hybrid $H_{%
\mathrm{V}}$, which allow us to avoid additional uncertainties in extracting
SR data for $g_{1}(q^{2})$. In the channel of the $\rho $ meson, we employ
\begin{equation}
M_{2}^{2}\in \lbrack 1,2]~\mathrm{GeV}^{2},\ s_{0}^{\prime }\in \lbrack
0.9,1.1]~\mathrm{GeV}^{2}.  \label{eq:Wind2}
\end{equation}

\begin{figure}[h]
\includegraphics[width=8.5cm]{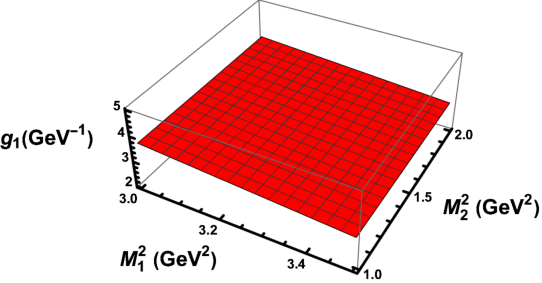}
\caption{The SR result for the form factor $g_1(Q^2)$ at $Q^{2}=10~\mathrm{%
GeV}^{2}$, $s_{0}=9~\mathrm{GeV}^{2}$ and $s_{0}^{\prime }=1~\mathrm{GeV}%
^{2} $ as a function of the Borel parameters $M_{1}^{2}$ and $M_{2}^{2}$. }
\label{fig:SR}
\end{figure}

The information about parameters permits us to carry out numerical
computation of the form factor $g_{1}(q^{2})$. At the pion's mass shell $%
q^{2}=m_{\pi }^{2}$ this function is equal to the strong coupling $g_{1}$
which is necessary to determine the partial width of the decay $H_{\mathrm{V}%
}\rightarrow \rho ^{-}\pi ^{+}$. But the SR method can be applied for
calculation of $g_{1}(q^{2})$ in the Euclidean region $q^{2}<0$. To avoid
this problem, we introduce the function $g_{1}(Q^{2})$ with $Q^{2}=-q^{2}$
and use it in our analysis. As an example, in Fig.\ \ref{fig:SR} we plot the
SR result obtained for $g_{1}(Q^{2})$ at $Q^{2}=10~\mathrm{GeV}^{2}$, $%
s_{0}=9~\mathrm{GeV}^{2}$ and $s_{0}^{\prime }=1~\mathrm{GeV}^{2}$ as a
function of the Borel parameters $M_{1}^{2}$ and $M_{2}^{2}$. One can see,
that there is residual dependence of $g_{1}(Q^{2})$ on these parameters. At
this point, we find%
\begin{equation}
g_{1}(Q^{2}=10~\mathrm{GeV}^{2})=(3.35\pm 0.65)\ \mathrm{GeV}^{-1},
\end{equation}%
where the theoretical uncertainties are equal to $\pm 19\%$ of the central
value and remain within limits acceptable for the SR computations.

\begin{widetext}

\begin{figure}[h!]
\begin{center}
\includegraphics[totalheight=6cm,width=8cm]{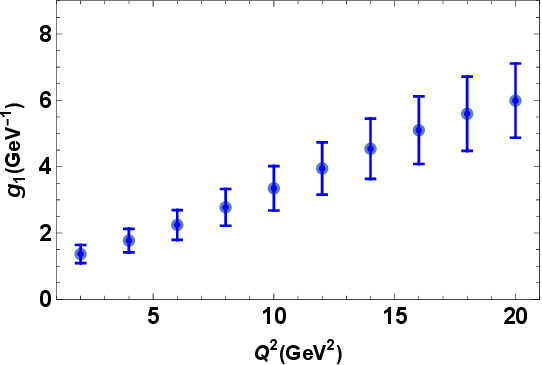}
\includegraphics[totalheight=6cm,width=8cm]{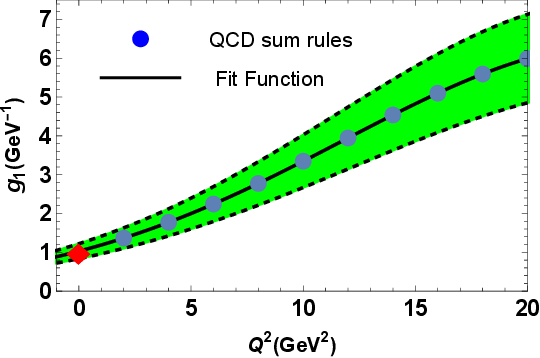}
\end{center}
\caption{The SR data for the form factor $g_1(Q^2)$ with corresponding error bars (left panel). The fit function $\mathcal{F}_1(Q^{2})$ (right panel). The red diamond marks the point $Q^{2}=-m_{\protect\pi}^{2}$, where the strong
coupling $g_1$ is evaluated.}
\label{fig:Fit}
\end{figure}

\end{widetext}

We perform the same analysis for $g_{1}(Q^{2})$ in the region $Q^{2}=2-20~%
\mathrm{GeV}^{2}$ depicted in the left panel of Fig.\ \ref{fig:Fit}.
Afterward, we use these SR data to find the extrapolating function $%
\mathcal{F}_{1}(Q^{2})$ which at the $Q^{2}=2-20~\mathrm{GeV}^{2}$ coincides
with them, but can be easily extended into a domain of positive $Q^{2}$. The
function
\begin{equation}
\mathcal{F}_{i}(Q^{2})=\mathcal{F}_{i}^{0}\mathrm{\exp }\left[ c_{i}^{1}%
\frac{Q^{2}}{m^{2}}+c_{i}^{2}\left( \frac{Q^{2}}{m^{2}}\right) ^{2}\right]
\label{eq:FitF}
\end{equation}%
is one of convenient choices for these purposes. Here, $\mathcal{F}_{i}^{0}$%
, $c_{i}^{1}$, and $c_{i}^{2}$ are parameters that are fixed from comparison
with the SR data. Then, it is not difficult to extract the constants $%
\mathcal{F}_{1}^{0}=1.03~\mathrm{GeV}^{-1}$, $c_{1}^{1}=0.78$, and $%
c_{1}^{2}=-0.08$. The function $\mathcal{F}_{1}(Q^{2})$ is also plotted in
Fig.\ \ref{fig:Fit} (right panel), in which is seen nice agreement between $%
\mathcal{F}_{1}(Q^{2})$ and the SR data.

For the strong coupling $g_{1}$, we find
\begin{equation}
g_{1}\equiv \mathcal{F}_{1}(-m_{\pi }^{2})=(1.03\pm 0.20)\ \mathrm{GeV}^{-1}.
\label{eq:Coupl1}
\end{equation}%
The width of the process $H_{\mathrm{V}}\rightarrow \rho ^{-}\pi ^{+}$ is
given by the formula
\begin{equation}
\Gamma \left[ H_{\mathrm{V}}\rightarrow \rho ^{-}\pi ^{+}\right] =g_{1}^{2}%
\frac{\lambda _{1}^{3}}{12\pi }.  \label{eq:PDw2}
\end{equation}%
Here,  $\lambda _{1}=\lambda (m,m_{\rho },m_{\pi })$ and the function $%
\lambda (x,y,z)$ has the form
\begin{equation}
\lambda (x,y,z)=\frac{\sqrt{%
x^{4}+y^{4}+z^{4}-2(x^{2}y^{2}+x^{2}z^{2}+y^{2}z^{2})}}{2x}.
\end{equation}%
As a result, we find
\begin{equation}
\Gamma \left[ H_{\mathrm{V}}\rightarrow \rho ^{-}\pi ^{+}\right] =(33.3\pm
10.1)~\mathrm{MeV}.  \label{eq:DW1}
\end{equation}%
The width $\Gamma \left[ H_{\mathrm{V}}\rightarrow \rho ^{+}\pi ^{-}\right] $
of the second decay is also equal to Eq.\ (\ref{eq:DW1}). \ Therefore, the
contribution of the processes $H_{\mathrm{V}}\rightarrow \rho \pi $ to the
full width of the hybrid meson $H_{\mathrm{V}}$ amounts to $\approx 67~%
\mathrm{MeV}$.

\section{Processes $H_{\mathrm{V}}\rightarrow b_{1}\protect\pi ,\ f_{1}%
\protect\pi $, and $f_{1}^{\prime }\protect\pi $}

\label{sec:Decays2}

This section is devoted to analysis of the decays $H_{\mathrm{V}}\rightarrow
b_{1}\pi $, $f_{1}(1285)\pi $, and $f_{1}(1420)\pi $. The $b_{1}(1235)$ is
the axial-vector meson with spin-parity $J^{\mathrm{PC}}={1^{+-}}$, whereas $%
f_{1}(1285)$ and $f_{1}(1420)$ are mesons with the quantum numbers ${1}^{++}$%
. In the case of the $b_{1}$ meson, we are going to study processes $H_{%
\mathrm{V}}\rightarrow b_{1}^{-}\pi ^{+}$ and $H_{\mathrm{V}}\rightarrow
b_{1}^{+}\pi ^{-\text{ }}$. The mesons $f_{1}$ and $f_{1}^{\prime }$ appear
in the decays $H_{\mathrm{V}}\rightarrow $ $f_{1}\pi $ and $H_{\mathrm{V}%
}\rightarrow $ $f_{1}^{\prime }\pi $.


\subsection{Decays $H_{\mathrm{V}}\rightarrow b_{1}^{-}\protect\pi ^{+}$ and
$H_{\mathrm{V}}\rightarrow b_{1}^{+}\protect\pi ^{-}$}


The partial widths of these processes are equal, therefore we investigate
only one of them. Let us analyze the decay $H_{\mathrm{V}}\rightarrow
b_{1}^{-}\pi ^{+}$. To determine the coupling $g_{2}$ responsible for the
strong interaction of particles at the vertex $H_{\mathrm{V}}b_{1}^{-}\pi
^{+}$, we start from the correlation function
\begin{eqnarray}
\widetilde{\Pi }_{\mu \nu }(p,p^{\prime }) &=&i^{2}\int
d^{4}xd^{4}ye^{ip^{\prime }x+iyq}\langle 0|\mathcal{T}\{J_{\nu }^{b}(x)
\notag \\
&&\times J^{\pi }(y)J_{\mu }^{\dagger }(0)\}|0\rangle .  \label{eq:CF2}
\end{eqnarray}%
The currents for the hybrid state $J_{\mu }(x)$ and pion $J^{\pi }(x)$ have
been defined above. The interpolating current for the meson $b_{1}^{-}$ is
given by the formula
\begin{equation}
J_{\nu }^{b}(x)=\overline{u}(x)\overleftrightarrow{D}_{\nu }(x)\gamma
_{5}d(x),  \label{eq:C2}
\end{equation}%
where $\overleftrightarrow{D}_{\nu }$ is
\begin{equation}
\overleftrightarrow{D}_{\nu }(x)=\frac{1}{2}\left[ \overrightarrow{D}(x)-%
\overleftarrow{D}(x)\right] .  \label{eq:Der}
\end{equation}%
Here,
\begin{eqnarray}
\overrightarrow{D}_{\nu }(x) &=&\overrightarrow{\partial }_{\nu }-i\frac{%
g_{s}}{2}\lambda ^{n}A_{\nu }^{n}(x),  \notag \\
\overleftarrow{D}_{\nu }(x) &=&\overleftarrow{\partial }_{\nu }+i\frac{g_{s}%
}{2}\lambda ^{n}A_{\nu }^{n}(x),
\end{eqnarray}%
where $A_{\nu }^{a}(x)$ is the external gluon field. In the Fock-Schwinger
gauge $x^{\nu }A_{\nu }^{a}(x)=0$ this field can be expressed in terms of
the gluon field strength tensor%
\begin{equation}
A_{\nu }^{n}(x)=\frac{1}{2}x^{\beta }G_{\beta \nu }^{n}(0)+\frac{1}{3}%
x^{\alpha }x^{\beta }D_{\alpha }G_{\beta \nu }^{n}(0)+....
\end{equation}

The sum rule for the form factor $g_{2}(q^{2})$ is extracted from analysis
of the correlator $\widetilde{\Pi }_{\mu \nu }(p,p^{\prime })$. Then, the
first component of the SR equality reads
\begin{eqnarray}
&&\widetilde{\Pi }_{\mu \nu }^{\mathrm{Phys}}(p,p^{\prime })=\frac{\langle
0|J_{\nu }^{b}|b_{1}^{-}(p^{\prime },\varepsilon ^{\prime })\rangle \langle
0|J^{\pi }|\pi ^{+}(q)\rangle }{(p^{\prime 2}-m_{b}^{2})(q^{2}-m_{\pi }^{2})}
\notag \\
&&\times \frac{\langle b_{1}^{-}(p^{\prime },\varepsilon ^{\prime })\pi
^{+}(q)|H_{\mathrm{V}}(p,\varepsilon )\rangle \langle H_{\mathrm{V}%
}(p,\varepsilon )|J_{\mu }^{\dagger }|0\rangle }{(p^{2}-m^{2})},
\label{eq:CorrF3}
\end{eqnarray}%
with $m_{b}=(1229.5\pm 3.2)~\mathrm{MeV}$ being the mass of the meson $%
b_{1}^{-}$. To recast $\widetilde{\Pi }_{\mu \nu }^{\mathrm{Phys}%
}(p,p^{\prime })$ into a form suitable for further analysis, we make use of
the following new matrix element
\begin{equation}
\langle 0|J_{\nu }^{b}|b_{1}^{-}(p^{\prime },\varepsilon ^{\prime })\rangle
=m_{b}f_{b}\varepsilon _{\nu }^{\prime },\   \label{eq:ME3}
\end{equation}%
where $f_{b}=181~\mathrm{MeV}$ is the decay constant of $b_{1}^{-}$. The
gauge-invariant expression for the matrix element $\langle
b_{1}^{-}(p^{\prime },\varepsilon ^{\prime })\pi ^{+}(q)|H_{\mathrm{V}%
}(p,\varepsilon )\rangle $ in the case of the
vector-axial-vector-pseudoscalar particles is
\begin{eqnarray}
&&\langle b_{1}^{-}(p^{\prime },\varepsilon ^{\prime })\pi ^{+}(q)|H_{%
\mathrm{V}}(p,\varepsilon )\rangle =g_{2}(q^{2})\left[ (p\cdot
q)(\varepsilon \cdot \varepsilon ^{\prime \ast })\right.  \notag \\
&&\left. -(p\cdot \varepsilon ^{\prime \ast })(q\cdot \varepsilon )\right] .
\label{eq:Ver2}
\end{eqnarray}

This new and known matrix elements of the particles $H_{\mathrm{V}}$ and $%
\pi ^{+}$ allow us to rewrite $\widetilde{\Pi }_{\mu \nu }^{\mathrm{Phys}%
}(p,p^{\prime })$ in the following form%
\begin{eqnarray}
&&\widetilde{\Pi }_{\mu \nu }^{\mathrm{Phys}}(p,p^{\prime })=g_{2}(q^{2})%
\frac{mfm_{b}f_{b}f_{\pi }m_{\pi }^{2}}{2m_{q}(p^{\prime
2}-m_{b}^{2})(p^{2}-m^{2})}  \notag \\
&&\times \frac{1}{(q^{2}-m_{\pi }^{2})}\left[ \frac{m^{2}-m_{b}^{2}+q^{2}}{2}%
g_{\mu \nu }-p_{\mu }p_{\nu }\right.  \notag \\
&&\left. +p_{\mu }^{\prime }p_{\nu }-\frac{m^{2}}{m_{b}^{2}}p_{\mu }^{\prime
}p_{\nu }^{\prime }+\frac{m^{2}+m_{b}^{2}-q^{2}}{2m_{b}^{2}}p_{\mu }p_{\nu
}^{\prime }\right] .  \label{eq:Phys2}
\end{eqnarray}%
The QCD side of the required equality reads%
\begin{eqnarray}
&&\widetilde{\Pi }_{\mu \nu }^{\mathrm{OPE}}(p,p^{\prime })=\frac{%
g_{s}\lambda _{ab}^{n}}{2\sqrt{2}}i\int d^{4}xd^{4}ye^{ip^{\prime
}x+iyq}G_{\mu \theta }^{n}(0)  \notag \\
&&\times \left\{ \mathrm{Tr}\left[ \gamma _{5}S_{d}^{ij}(x-y)\gamma
_{5}S_{u}^{jb}(y)\gamma ^{\theta }\overleftrightarrow{D}_{\nu
}(x)S_{u}^{ai}(-x)\right] \right.  \notag \\
&&\left. -\mathrm{Tr}\left[ \gamma _{5}S_{d}^{ib}(x)\gamma ^{\theta
}S_{d}^{aj}(-y)\gamma _{5}\overleftrightarrow{D}_{\nu }(x)S_{u}^{ji}(y-x)%
\right] \right\} .  \notag \\
&&  \label{eq:OPE2}
\end{eqnarray}

\begin{figure}[h]
\includegraphics[width=8.5cm]{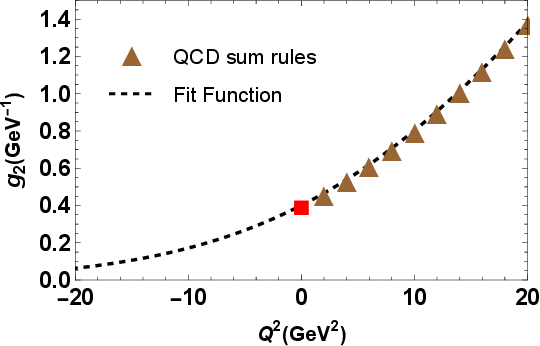}
\caption{The sum rule's data and the extrapolating function $\mathcal{F}%
_2(Q^{2})$. The red square shows the position $Q^{2}=-m_{\protect\pi}^{2}$.}
\label{fig:Fit1}
\end{figure}

The SR for the form factor $g_{2}(q^{2})$ is obtained using the invariant
amplitudes $\widetilde{\Pi }^{\mathrm{Phys}}\left( p^{2},p^{\prime
2},q^{2}\right) $ and $\widetilde{\Pi }^{\mathrm{OPE}}(p^{2},p^{\prime
2},q^{2})$ that correspond to terms $p_{\mu }p_{\nu }$ both in $\widetilde{%
\Pi }_{\mu \nu }^{\mathrm{Phys}}(p,p^{\prime })$ and $\widetilde{\Pi }_{\mu
\nu }^{\mathrm{OPE}}(p,p^{\prime })$. After standard Borel transformations
and continuum subtraction, one gets
\begin{eqnarray}
&&g_{2}(q^{2})=\frac{2m_{q}(q^{2}-m_{\pi }^{2})}{mfm_{b}f_{b}f_{\pi }m_{\pi
}^{2}}e^{m^{2}/M_{1}^{2}}e^{m_{b}^{2}/M_{2}^{2}}  \notag \\
&&\times \widetilde{\Pi }(\mathbf{M}^{2},\mathbf{s}_{0},q^{2}).
\label{eq:SRg2}
\end{eqnarray}%
The remaining operations are similar to ones explained in the previous
section, therefore we omit these details. The difference here is connected
with the choice of the region for the parameters in the $b_{1}^{-}$ meson
channel: It is introduced in the form
\begin{equation}
M_{2}^{2}\in \lbrack 1.5,2.5]~\mathrm{GeV}^{2},\ s_{0}^{\prime }\in \lbrack
2,2.5]~\mathrm{GeV}^{2}.  \label{eq:Wind3}
\end{equation}%
The results of the SR computations are shown in Fig.\ \ref{fig:Fit1}. The
fit function $\mathcal{F}_{2}(Q^{2})$ employed to calculate the coupling $%
g_{2}$ is fixed due to the parameters%
\begin{equation}
\mathcal{F}_{2}^{0}=0.40~\mathrm{GeV}^{-1},c_{2}^{1}=0.41,c_{2}^{2}=-0.02.
\end{equation}%
Then, it is easy to evaluate $g_{2}$%
\begin{equation}
g_{2}\equiv \mathcal{F}_{2}(-m_{\pi }^{2})=(4.0\pm 0.8)\times 10^{-1}\
\mathrm{GeV}^{-1}.
\end{equation}%
The width of the process $H_{\mathrm{V}}\rightarrow b_{1}^{-}\pi ^{+}$ is
given by the expression
\begin{equation}
\Gamma \left[ H_{\mathrm{V}}\rightarrow b_{1}^{-}\pi ^{+}\right] =g_{2}^{2}%
\frac{\lambda _{2}}{24\pi m^{2}}|M|^{2}.
\end{equation}%
Here,
\begin{eqnarray}
|M|^{2} &=&\frac{1}{4m_{b}^{2}}\left[ m^{6}-2m_{\pi
}^{2}m^{4}+2m_{b}^{2}(m_{b}^{2}-m_{\pi }^{2})^{2}\right.  \notag \\
&&\left. +m^{2}(m_{\pi }^{4}-3m_{b}^{4}+6m_{b}^{2}m_{\pi }^{2})\right] ,
\end{eqnarray}%
and $\lambda _{2}=\lambda _{2}(m,m_{b},m_{\pi }).$ The partial width of this
decay is equal to
\begin{equation}
\Gamma \left[ H_{\mathrm{V}}\rightarrow b_{1}^{-}\pi ^{+}\right] =(6.4\pm
1.8)~\mathrm{MeV}.  \label{eq:DW2}
\end{equation}%
The width of the second process $H_{\mathrm{V}}\rightarrow b_{1}^{+}\pi ^{-}$
is also given by Eq.\ (\ref{eq:DW2}).

\subsection{Processes $H_{\mathrm{V}}\rightarrow f_{1}\protect\pi $ and $%
f_{1}^{\prime }\protect\pi $}

The treatment of the decays $H_{\mathrm{V}}\rightarrow f_{1}\pi $ and $H_{%
\mathrm{V}}\rightarrow f_{1}^{\prime }\pi $ due to mixing in the system $%
f_{1}-f_{1}^{\prime }$ differs from analyses of the previous processes.
Because the mesons $f_{1}$ and $f_{1}^{\prime }$ have the nonstrange and
strange components, it is convenient to consider them in the quark-flavor
basis. Originally, this basis was introduced to treat the mixing in the $%
\eta -\eta ^{\prime }$ system \cite{Feldmann:1998vh}, and later was used to
explore different processes with $\eta $ and $\eta ^{\prime }$ mesons \cite%
{Agaev:2014wna,Agaev:2015faa}. But it can also be applied to the system of
the axial-vector mesons $f_{1}$ and $f_{1}^{\prime }$ as well \cite%
{Jiang:2020eml}.

In this scheme the physical mesons $f_{1}$ and $f_{1}^{\prime }$ can be
presented using the basic states $|f_{1q}\rangle =(\overline{u}u+\overline{d}%
d)/\sqrt{2}$ and $|f_{1s}\rangle =\overline{s}s$ through the formula%
\begin{equation}
\begin{pmatrix}
f_{1} \\
f_{1}^{\prime }%
\end{pmatrix}%
=U(\phi )%
\begin{pmatrix}
|f_{1q}\rangle \\
|f_{1s}\rangle%
\end{pmatrix}%
.
\end{equation}%
Here
\begin{equation}
U(\phi )=%
\begin{pmatrix}
\cos \phi & -\sin \phi \\
\sin \phi & \cos \phi%
\end{pmatrix}%
,
\end{equation}%
is the mixing matrix and $\phi $ is the mixing angle.

This assumption on the state mixing implies that the same pattern also
applies to relevant currents, decay constants and wave functions. Then, in
the quark-flavor basis the axial-vector currents for the mesons $f_{1}$ and $%
f_{1}^{\prime }$ acquire the following forms
\begin{equation}
\begin{pmatrix}
J_{\nu }^{f_{1}}(x) \\
J_{\nu }^{f_{1}^{\prime }}(x)%
\end{pmatrix}%
=U(\phi )%
\begin{pmatrix}
J_{\nu }^{q}(x) \\
J_{\nu }^{s}(x)%
\end{pmatrix}%
,
\end{equation}%
where
\begin{eqnarray}
J_{\nu }^{q}(x) &=&\frac{1}{\sqrt{2}}\left[ \overline{u}_{i}(x)\gamma _{\nu
}\gamma _{5}u_{i}(x)+\overline{d}_{i}(x)i\gamma _{5}d_{i}(x)\right] ,  \notag
\\
J_{\nu }^{s}(x) &=&\overline{s}_{i}(x)\gamma _{\nu }\gamma _{5}s_{i}(x).
\end{eqnarray}%
As is seen, the currents $J_{\nu }^{f_{1}}(x)$ and $J_{\nu }^{f_{1}^{\prime
}}(x)$ have the $\overline{q}q$ and $\overline{s}s$ components. It is also
clear that only $\overline{q}q$ components of the mesons $f_{1}$ and $%
f_{1}^{\prime }$ contribute to the decays under consideration.

The correlation function which should be explored is given by the formula%
\begin{eqnarray}
\widehat{\Pi }_{\mu \nu }(p,p^{\prime }) &=&i^{2}\int
d^{4}xd^{4}ye^{ip^{\prime }x+iyq}\langle 0|\mathcal{T}\{J_{\nu }^{f_{1}}(x)
\notag \\
&&\times J^{\pi 0}(y)J_{\mu }^{\dagger }(0)\}|0\rangle ,  \label{eq:CF3}
\end{eqnarray}%
where $J_{\nu }^{f_{1}}(x)$ and $J^{\pi 0}(x)$ are interpolating currents of
the mesons $f_{1}$ and $\pi ^{0}$. They are defined by the expressions%
\begin{equation}
J_{\nu }^{f_{1}}(x)=\cos \phi J_{\nu }^{q}(x),  \label{eq:CR3}
\end{equation}%
and
\begin{eqnarray}
J^{\pi 0}(x) &=&\frac{1}{\sqrt{2}}[\overline{u}_{j}(x)i\gamma _{5}u_{j}(x)-%
\overline{d}_{j}(x)i\gamma _{5}d_{j}(x)].  \notag \\
&&  \label{eqCR4}
\end{eqnarray}%
For the correlation function $\widehat{\Pi }_{\mu \nu }^{\mathrm{Phys}%
}(p,p^{\prime })$, we get
\begin{eqnarray}
&&\widehat{\Pi }_{\mu \nu }^{\mathrm{Phys}}(p,p^{\prime })=\frac{\langle
0|J_{\nu }^{f_{1}}|f_{1}(p^{\prime },\varepsilon ^{\prime })\rangle \langle
0|J^{\pi 0}|\pi ^{0}(q)\rangle }{(p^{\prime 2}-m_{f_{1}}^{2})(q^{2}-m_{\pi
}^{2})}  \notag \\
&&\times \frac{\langle f_{1}(p^{\prime },\varepsilon ^{\prime })\pi
^{0}(q)|H_{\mathrm{V}}(p,\varepsilon )\rangle \langle H_{\mathrm{V}%
}(p,\varepsilon )|J_{\mu }^{\dagger }|0\rangle }{(p^{2}-m^{2})},
\end{eqnarray}%
where $m_{f_{1}}=(1281.8\pm 0.5)~\mathrm{MeV}$ and $\varepsilon _{\nu
}^{\prime }$ are the mass and polarization vector of the meson $f_{1}$. To
find $\langle 0|J_{\nu }^{f_{1}}|f_{1}(p^{\prime },\varepsilon ^{\prime
})\rangle $ we introduce the matrix elements
\begin{eqnarray}
\langle 0|J_{\nu }^{q}|f_{1}(p^{\prime },\varepsilon ^{\prime })\rangle
&=&m_{f_{1}}f_{f_{1}}^{q}\varepsilon _{\nu }^{\prime },  \notag \\
\langle 0|J_{\nu }^{s}|f_{1}(p^{\prime },\varepsilon ^{\prime })\rangle
&=&m_{f_{1}}f_{f_{1}}^{s}\varepsilon _{\nu }^{\prime },
\end{eqnarray}%
where the decay constants $f_{f_{1}}^{q}$ and $f_{f_{1}}^{s}$ can be
determined by means of the formula%
\begin{equation}
\begin{pmatrix}
f_{f_{1}}^{q} & f_{f_{1}}^{s} \\
f_{f_{1}^{\prime }}^{q} & f_{f_{1}^{\prime }}^{s}%
\end{pmatrix}%
=U(\phi )%
\begin{pmatrix}
f_{1q} & 0 \\
0 & f_{1s}%
\end{pmatrix}%
.
\end{equation}%
In other words, all four decay constants are expressed in terms of the two
parameters $f_{1q}$ and $f_{1s}$.

The vertex $\langle f_{1}(p^{\prime },\varepsilon ^{\prime })\pi ^{0}(q)|H_{%
\mathrm{V}}(p,\varepsilon )\rangle $ has the following form%
\begin{eqnarray}
&&\langle f_{1}(p^{\prime },\varepsilon ^{\prime })\pi ^{0}(q)|H_{\mathrm{V}%
}(p,\varepsilon )\rangle =g_{3}(q^{2})\left[ (p\cdot q)(\varepsilon \cdot
\varepsilon ^{\prime \ast })\right.  \notag \\
&&\left. -(p\cdot \varepsilon ^{\prime \ast })(q\cdot \varepsilon )\right] .
\end{eqnarray}%
Then it is not difficult to write down
\begin{eqnarray}
&&\widehat{\Pi }_{\mu \nu }^{\mathrm{Phys}}(p,p^{\prime })=\frac{mfm_{\pi
}^{2}m_{f_{1}}f_{1q}\cos ^{2}\phi }{2m_{q}(p^{\prime
2}-m_{f_{1}}^{2})(p^{2}-m^{2})}  \notag \\
&&\times \frac{1}{(q^{2}-m_{\pi }^{2})}\left[ \frac{m^{2}-m_{f_{1}}^{2}+q^{2}%
}{2}g_{\mu \nu }-p_{\mu }p_{\nu }\right.  \notag \\
&&\left. +p_{\mu }^{\prime }p_{\nu }-\frac{m^{2}}{m_{f_{1}}^{2}}p_{\mu
}^{\prime }p_{\nu }^{\prime }+\frac{m^{2}+m_{f_{1}}^{2}-q^{2}}{2m_{f_{1}}^{2}%
}p_{\mu }p_{\nu }^{\prime }\right] .
\end{eqnarray}

The correlator $\widehat{\Pi }_{\mu \nu }^{\mathrm{OPE}}(p,p^{\prime })$ is
given by the expression
\begin{eqnarray}
&&\widehat{\Pi }_{\mu \nu }^{\mathrm{OPE}}(p,p^{\prime })=i\frac{%
g_{s}\lambda _{ab}^{n}\cos \phi }{2\sqrt{2}}\int d^{4}xd^{4}ye^{ip^{\prime
}x+iyq}G_{\mu \theta }^{n}(0)  \notag \\
&&\times \left\{ \mathrm{Tr}\left[ \gamma _{\nu }\gamma
_{5}S_{q}^{ib}(x)\gamma ^{\theta }S_{q}^{aj}(-y)\gamma _{5}S_{q}^{ji}(y-x)%
\right] \right.  \notag \\
&&\left. +\mathrm{Tr}\left[ \gamma _{\nu }\gamma _{5}S_{q}^{ij}(x-y)\gamma
_{5}S_{q}^{jb}(y)\gamma ^{\theta }S_{q}^{ai}(-x)\right] \right\} ,
\label{eq:OPE4}
\end{eqnarray}%
where $S_{q}(x)$ is the light $d$ or $u$ propagator. The sum rule for the
form factor $g_{3}(q^{2})$ reads
\begin{eqnarray}
&&g_{3}(q^{2})=\frac{2m_{q}(q^{2}-m_{\pi }^{2})}{mfm_{\pi
}^{2}m_{f_{1}}f_{1q}\cos \phi }e^{m^{2}/M_{1}^{2}}e^{m_{f_{1}}^{2}/M_{2}^{2}}
\notag \\
&&\times \widehat{\Pi }(\mathbf{M}^{2},\mathbf{s}_{0},q^{2}).
\label{eq:SRg3}
\end{eqnarray}%
Let us note that to extract the sum rule Eq.\ (\ref{eq:SRg3}) we have used
invariant amplitudes $\widehat{\Pi }^{\mathrm{Phys}}\left( p^{2},p^{\prime
2},q^{2}\right) $ and $\widehat{\Pi }^{\mathrm{OPE}}(p^{2},p^{\prime
2},q^{2})$ that correspond to terms $p_{\mu }p_{\nu }$ in the physical and $%
\mathrm{OPE}$ expressions of the correlator $\widehat{\Pi }_{\mu \nu
}(p,p^{\prime })$.

The numerical analysis are carried out after fixing the parameters of the $%
f_{1}-f_{1}^{\prime }$ system. For the mixing angle $\phi $ and decay
constants $f_{_{1}q}$ and $f_{1s}$ we employ the values
\begin{eqnarray}
\phi &=&(24.0_{-2.7}^{+3.2})^{\circ },\ f_{1q}=193_{-38}^{+43}\ \mathrm{MeV,}
\notag \\
f_{1s} &=&(230\pm 9)\ \mathrm{MeV}.  \label{eq:DCf}
\end{eqnarray}%
The mixing angle $\phi $ in Eq.\ (\ref{eq:DCf}) was extracted by the LHCb
collaboration from the process $B_{d,s}^{0}\rightarrow J/\psi f_{1}(1285)$
\cite{LHCb:2013ged}. The decay constants $f_{1q}$ and $f_{1s}$ were
estimated in the framework of the covariant light-front quark model in Ref.\
\cite{Verma:2011yw}.

\begin{figure}[h]
\includegraphics[width=8.5cm]{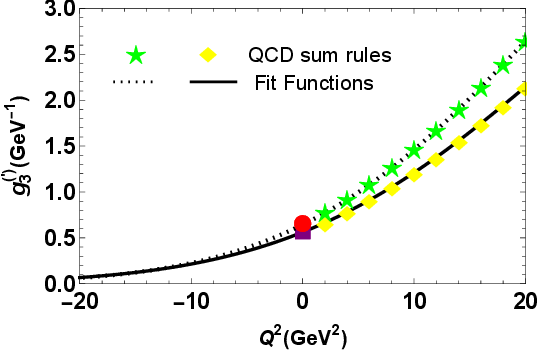}
\caption{QCD data and extrapolating functions $\mathcal{F}_{3}(Q^{2})$
(solid line) and $\mathcal{F}_{3}^{\prime}(Q^{2})$ (dotted line) The circle
and square fix the point $Q^{2}=-m_{\protect\pi }^{2}$.}
\label{fig:Fit2}
\end{figure}

Afterward, we apply the operations which have been explained above.
Therefore, we omit details and write down only principal results. They have
been found by employing the following Borel and continuum subtraction
parameters: In the of the hybrid $H_{\mathrm{V}}$ we have used the regions
Eq.\ (\ref{eq:Wind1}), whereas for the $f_{1}$ channel employed
\begin{equation}
M_{2}^{2}\in \lbrack 1.5,2.5]~\mathrm{GeV}^{2},\ s_{0}^{\prime }\in \lbrack
1.8,2.0]~\mathrm{GeV}^{2}.
\end{equation}%
Then the strong coupling $g_{3}$ computed at the mass shell $Q^{2}=-m_{\pi
}^{2}$ amounts to
\begin{equation}
g_{3}\equiv \mathcal{F}_{3}(-m_{\pi }^{2})=(5.6\pm 1.3)\times 10^{-1}\
\mathrm{GeV}^{-1},
\end{equation}%
where $\mathcal{F}_{3}(Q^{2})$ is the fitting function with parameters $%
\mathcal{F}_{3}^{0}=0.56~\mathrm{GeV}^{-1},c_{3}^{1}=0.46,c_{3}^{2}=-0.03$
(see, Fig.\ \ref{fig:Fit2}). The partial width of the decay $H_{\mathrm{V}%
}\rightarrow f_{1}\pi $ is%
\begin{equation}
\Gamma \left[ H_{\mathrm{V}}\rightarrow f_{1}\pi \right] =(10.7\pm 3.6)~%
\mathrm{MeV}.
\end{equation}

The decay $H_{\mathrm{V}}\rightarrow f_{1}^{\prime }\pi $ can be studied in
the same manner. The difference here is connected with $\sin \phi $ and $%
m_{f_{1}^{\prime }}=1428.4_{-1.3}^{+1.5}~\mathrm{MeV}$ in the relevant
formulas. As a result, we find
\begin{equation}
g_{3}^{\prime }\equiv \mathcal{F}_{3}^{\prime }(-m_{\pi }^{2})=(6.7\pm
1.4)\times 10^{-1}\ \mathrm{GeV}^{-1}.
\end{equation}%
In these calculations in the $f_{1}^{\prime }$ channel we have employed the
regions

\begin{equation}
M_{2}^{2}\in \lbrack 1.5,2.5]~\mathrm{GeV}^{2},\ s_{0}^{\prime }\in \lbrack
2.9,3.0]~\mathrm{GeV}^{2}.
\end{equation}%
The width of the decay $H_{\mathrm{V}}\rightarrow f_{1}^{\prime }\pi $
amounts to
\begin{equation}
\Gamma \left[ H_{\mathrm{V}}\rightarrow f_{1}^{\prime }\pi \right] =(9.6\pm
2.9)~\mathrm{MeV}.
\end{equation}

\section{Reactions $H_{\mathrm{V}}\to \protect\eta \protect\pi $ and $H_{%
\mathrm{V}}\to \protect\eta^{\prime} \protect\pi $}

\label{sec:Decays3}
The reactions $H_{\mathrm{V}}\rightarrow \eta \pi $ and $H_{\mathrm{V}%
}\rightarrow \eta ^{\prime }\pi $ can be investigated by employing the
technical tools described in the previous section. Here, because of $U(1)$
anomaly we have to take into account the mixing in the $\eta -\eta ^{\prime
} $ system. The physical mesons $\eta $ and $\eta ^{\prime }$, in general,
can be formed using either the octet-singlet or quark-flavor bases of the
flavor $SU_{f}(3)$ group \cite{Feldmann:1998vh}. Mixing of the physical
states, decay constants, leading and higher twist distribution amplitudes in
the quark-flavor basis $|\eta _{q}\rangle =(\overline{u}u+\overline{d}d)/%
\sqrt{2} $ and $|\eta _{s}\rangle =\overline{s}s$ have very simple form.

In the quark-flavor basis the physical mesons $\eta $ and $\eta ^{\prime }$
are expressed in terms of the basic states $|\eta _{q}\rangle $ and $|\eta
_{s}\rangle $ through the formula%
\begin{equation}
\begin{pmatrix}
\eta \\
\eta ^{\prime }%
\end{pmatrix}%
=U(\varphi )%
\begin{pmatrix}
|\eta _{q}\rangle \\
|\eta _{s}\rangle%
\end{pmatrix}%
.  \label{eq:MixPhys}
\end{equation}%
In Eq.\ (\ref{eq:MixPhys}) $U(\varphi )$ is the mixing matrix%
\begin{equation}
U(\varphi )=%
\begin{pmatrix}
\cos \varphi & -\sin \varphi \\
\sin \varphi & \cos \varphi%
\end{pmatrix}%
,  \label{eq:Mixmat}
\end{equation}%
and $\varphi $ is the corresponding mixing angle. Then, in the quark-flavor
basis the currents for the pseudoscalar mesons $\eta $ and $\eta ^{\prime }$
acquire the following forms
\begin{equation}
\begin{pmatrix}
J^{\eta }(x) \\
J^{\eta ^{\prime }}(x)%
\end{pmatrix}%
=U(\varphi )%
\begin{pmatrix}
J_{q}(x) \\
J_{s}(x)%
\end{pmatrix}%
,  \label{eq:MixCurr}
\end{equation}%
where
\begin{eqnarray}
J_{q}(x) &=&\frac{1}{\sqrt{2}}\left[ \overline{u}_{i}(x)i\gamma _{5}u_{i}(x)+%
\overline{d}_{i}(x)i\gamma _{5}d_{i}(x)\right] ,  \notag \\
J_{s}(x) &=&\overline{s}_{i}(x)i\gamma _{5}s_{i}(x).
\end{eqnarray}

We start our studies from the process $H_{\mathrm{V}}\rightarrow \eta \pi $.
In this case, we have to investigate the correlation function
\begin{eqnarray}
\Pi _{\mu }(p,p^{\prime }) &=&i^{2}\int d^{4}xd^{4}ye^{ip^{\prime
}x+iyq}\langle 0|\mathcal{T}\{J^{\eta }(x)(x)  \notag \\
&&\times J^{\pi 0}(y)J_{\mu }^{\dagger }(0)\}|0\rangle .  \label{eq:CF4}
\end{eqnarray}%
The contribution of the ground-level particles to the correlation function $%
\Pi ^{\prime }(p,p^{\prime })$ has the form
\begin{eqnarray}
&&\Pi _{\mu }^{\mathrm{Phys}}(p,p^{\prime })=\frac{\langle 0|J^{\eta }|\eta
(p^{\prime })\rangle \langle 0|J^{\pi 0}|\pi ^{0}(q)\rangle }{%
(p^{2}-m^{2})(p^{\prime 2}-m_{\eta }^{2})}  \notag \\
&&\times \frac{\langle \pi ^{0}(q)\eta (p^{\prime })|H_{\mathrm{V}%
}(p,\varepsilon )\rangle \langle H_{\mathrm{V}}(p,\varepsilon )|J_{\mu
}^{\dagger }|0\rangle }{(q^{2}-m_{\pi }^{2})}  \notag \\
&&+\cdots ,  \label{eq:CF5}
\end{eqnarray}%
where the dots indicate effects of higher resonances and continuum states.
There are two new matrix elements in Eq.\ (\ref{eq:CF5}) which should be
defined before computation of $\Pi ^{\mathrm{Phys}}(p,p^{\prime })$. To fix
the first of them $\langle 0|J^{\eta }|\eta (p^{\prime })\rangle $ it is
convenient to introduce the matrix elements of the currents $J_{q}(x)$ and $%
J_{s}(x)$
\begin{eqnarray}
\ 2m_{q}\langle 0|J_{q}|\eta (p^{\prime })\rangle &=&h_{\eta }^{q},\   \notag
\\
2m_{s}\langle 0|J_{s}|\eta (p^{\prime })\rangle &=&h_{\eta }^{s},
\label{eq:Mel3}
\end{eqnarray}%
with $2m_{q}$ being equal to $m_{u}+m_{d}$. In the $|\eta _{q}\rangle -$ $%
|\eta _{s}\rangle $ basis the twist-3 matrix elements $h_{\eta }^{q}$ and $%
h_{\eta }^{s}$ is obtained from the expression%
\begin{equation}
\begin{pmatrix}
h_{\eta }^{q} & h_{\eta }^{s} \\
h_{\eta ^{\prime }}^{q} & h_{\eta ^{\prime }}^{s}%
\end{pmatrix}%
=U(\varphi )%
\begin{pmatrix}
h_{q} & 0 \\
0 & h_{s}%
\end{pmatrix}%
,  \label{eq:Dconst}
\end{equation}%
where $h_{s}=0.086~\mathrm{GeV}^{3}$, and $\varphi =39.3^{\circ }\pm
1.0^{\circ }$ were determined from a fit to experimental data \cite%
{Feldmann:1998vh} and $h_{q}=0.0025~\mathrm{GeV}^{3}$ was found from
theoretical analysis \cite{Kaiser:2000gs,Beneke:2002jn}. The next matrix
element $\langle \pi ^{0}(q)\eta (p^{\prime })|H_{\mathrm{V}}(p,\varepsilon
)\rangle $ is%
\begin{equation}
\langle \pi ^{0}(q)\eta (p^{\prime })|H_{\mathrm{V}}(p,\varepsilon )\rangle
=g_{4}(q^{2})\varepsilon \cdot p^{\prime }.  \label{eqME4}
\end{equation}

Having employed this input information, we get
\begin{eqnarray}
&&\Pi _{\mu }^{\mathrm{Phys}}(p,p^{\prime })=\frac{mff_{\pi }m_{\pi
}^{2}h_{q}\cos ^{2}\varphi }{4m_{q}^{2}(p^{2}-m^{2})(p^{\prime 2}-m_{\eta
}^{2})(q^{2}-m_{\pi }^{2})}  \notag \\
&&\times \left[ \frac{m^{2}+m_{\eta }^{2}-q^{2}}{2m^{2}}p_{\mu }-p_{\mu
}^{\prime }\right] +\cdots .
\end{eqnarray}%
The correlator $\Pi _{\mu }(p,p^{\prime })$ amounts to
\begin{eqnarray}
&&\Pi _{\mu }^{\mathrm{OPE}}(p,p^{\prime })=-\frac{\cos \varphi g_{s}\lambda
_{ab}^{n}}{2\sqrt{2}}i\int d^{4}xd^{4}ye^{ip^{\prime }x+iyq}G_{\mu \theta
}^{n}(0)  \notag \\
&&\times \left\{ \mathrm{Tr}\left[ \gamma _{5}S_{q}^{ib}(x)\gamma ^{\theta
}S_{q}^{aj}(-y)\gamma _{5}S_{q}^{ji}(y-x)\right] \right.  \notag \\
&&\left. +\mathrm{Tr}\left[ \gamma _{5}S_{q}^{ij}(x-y)\gamma
_{5}S_{q}^{jb}(y)\gamma ^{\theta }S_{q}^{ai}(-x)\right] \right\} .
\end{eqnarray}%
The SR for the form factor $g_{4}(q^{2})$ can be obtained by means of
amplitudes that correspond to structures $p_{\mu }$ in the correlators $\Pi
_{\mu }^{\mathrm{Phys}}(p,p^{\prime })$ and $\Pi _{\mu }^{\mathrm{OPE}%
}(p,p^{\prime })$.

\begin{figure}[h]
\includegraphics[width=8.5cm]{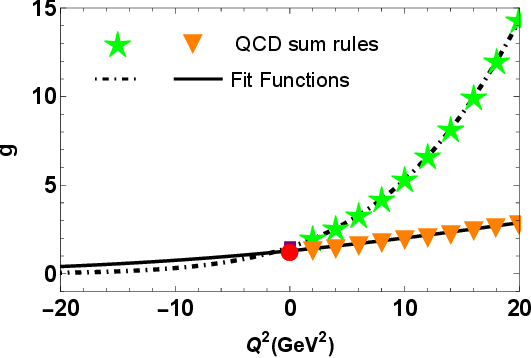}
\caption{The SR data and the extrapolating functions $\mathcal{F}_4(Q^{2})$
for the strong coupling $g_4$ (solid line) and $\mathcal{F}_5(Q^{2})$ for $%
g_5$ (dot-dashed line).}
\label{fig:Fit3}
\end{figure}

Numerical computations of $g_{4}(Q^{2})$ have been carried out at $%
Q^{2}=2-20~\mathrm{GeV}^{2}$, using the working windows%
\begin{equation}
M_{2}^{2}\in \lbrack 1,1.5]~\mathrm{GeV}^{2},\ s_{0}^{\prime }\in \lbrack
0.8,1]~\mathrm{GeV}^{2}.
\end{equation}%
The extrapolating function $\mathcal{F}_{4}(Q^{2})$ to fit the SR data is
determined by the parameters $\mathcal{F}_{4}^{0}=1.29$,$%
~c_{4}^{1}=0.26,c_{4}^{2}=-0.01.$ This function and corresponding SR data
are plotted in Fig.\ \ref{fig:Fit3}.

As a result, the coupling $g_{4}$ is%
\begin{equation}
g_{4}\equiv \mathcal{F}_{4}(-m_{\pi }^{2})=(1.29\pm 0.27).
\end{equation}%
The width of the process $H_{\mathrm{V}}\rightarrow \eta \pi $ is given by
the formula
\begin{equation}
\Gamma \left[ H_{\mathrm{V}}\rightarrow \eta \pi \right] =g_{4}^{2}\frac{%
\lambda _{4}^{3}}{24\pi m^{2}},
\end{equation}%
where $\lambda _{4}=\lambda (m,m_{\eta },m_{\pi })$. The partial width of
this channel is equal to
\begin{equation}
\Gamma \left[ H_{\mathrm{V}}\rightarrow \eta \pi \right] =(5.3\pm 1.6)~%
\mathrm{MeV}.
\end{equation}

The similar calculations are carried out for the $\eta ^{\prime }\pi $
final-state as well. Here, differences are connected with necessities to
replace in numerical analysis the factors $\cos \varphi \rightarrow $ $\sin
\varphi $, $m_{\eta }\rightarrow m_{\eta ^{\prime }}$ and also by the
regions for $M_{2}^{2}\in \lbrack 1.5,2]~\mathrm{GeV}^{2},\ s_{0}^{\prime
}\in \lbrack 1.3,1.7]~\mathrm{GeV}^{2}$. Our studies demonstrate that the
extrapolating function $\mathcal{F}_{5}(Q^{2})$ has the parameters $\mathcal{%
F}_{5}^{0}=1.49$,$~c_{5}^{1}=0.74,c_{5}^{2}=-0.04$ (see, Fig.\ \ref{fig:Fit3}%
). Then the strong coupling $g_{5}$ extracted from this function at $%
Q^{2}=-m_{\pi }^{2}$ is%
\begin{equation}
g_{5}\equiv \mathcal{F}_{5}(-m_{\pi }^{2})=(1.49\pm 0.32).
\end{equation}%
The partial width of the process $H_{\mathrm{V}}\rightarrow \eta ^{\prime
}\pi $ amounts to

\begin{equation}
\Gamma \left[ H_{\mathrm{V}}\rightarrow \eta ^{\prime }\pi \right] =(4.7\pm
1.4)~\mathrm{MeV}.
\end{equation}

Information gained in this and previous sections allows us to estimate the
full width of the hybrid $H_{\mathrm{V}}$ as
\begin{equation}
\Gamma \left[ H_{\mathrm{V}}\right] =(109.7\pm 16.0)~\mathrm{MeV}.
\end{equation}

\section{Discussion and conclusions}

\label{sec:Conc}

In the present article we have estimated the full width of the isovector
hybrid meson $H_{\mathrm{V}}$ with spin-parities $J^{\mathrm{PC}}={1^{-+}}$
and content $(\overline{u}gu-\overline{d}gd)/\sqrt{2}$. We have considered
the decays of this particle to $\rho \pi $, $b_{1}\pi $, $f_{1}\pi $, $%
f_{1}^{\prime }\pi $, $\eta \pi $, and $\eta ^{\prime }\pi $ mesons. The
partial width of these channels have been calculated using QCD three-point
sum rule method. This approach is necessary to evaluate the strong couplings
at the corresponding hybrid-meson-meson vertices.

The three-point sum rules for the form factors $g_{i}(q^{2})$ depend on the
two independent Borel parameters and contains double integral over
parameters $s_{0}$ and $s_{0}^{\prime }$. We have used SRs for numerical
computations of the QCD data which, by means of the fitting functions, have
been later extrapolated to regions of negative $Q^{2}$ to extract strong
couplings of interest. It is worth to emphasize that we have performed these
calculations without making any additional assumptions about the Borel
parameters or imposing limits on the particles' momenta. Our studies have
demonstrated that aforementioned decays contribute to the full width of the
hybrid $H_{\mathrm{V}}$ in the proportions: $\rho \pi $: $b_{1}\pi $: $%
f_{1}\pi $: $f_{1}^{\prime }\pi $: $\eta \pi $: $\eta ^{\prime }\pi
\rightarrow 61:12:10:9:5:3$. In other words, the dominant decay channel of $%
H_{\mathrm{V}}$ is the process $H_{\mathrm{V}}\rightarrow $ $\rho \pi $.
Important are also decays $H_{\mathrm{V}}\rightarrow $ $b_{1}\pi $ and $%
f_{1}^{(\prime )}\pi $. The result $\Gamma \left[ H_{\mathrm{V}}\right]
\approx 110~\mathrm{MeV}$ of the present article characterize the hybrid
meson $H_{\mathrm{V}}$ as a state with moderate full width.

The processes with $\eta \pi $ and $\eta ^{\prime }\pi $ mesons are
subdominant modes, but their partial widths are of the same order as ones of
the decays $H_{\mathrm{V}}\rightarrow $ $b_{1}\pi $, $f_{1}\pi $. The total
contribution of the decays $H_{\mathrm{V}}\rightarrow \eta \pi $, $\eta
^{\prime }\pi $ amounts to $9\%$ of the hybrid meson's full width $\Gamma %
\left[ H_{\mathrm{V}}\right] $. This is important result because in all
existing calculations, widths of these decays were found negligibly small.
The flux tube model even forbids decays to two ground-state mesons, i.e., to
$\eta \pi $, $\eta ^{\prime }\pi $ pairs. This conclusion contradicts to
experimental measurements, because structures $\pi _{1}(1400)$ and $\pi
_{1}(1600)$ were also seen in the $\eta \pi $ and $\eta ^{\prime }\pi $
channels \cite{Chen:2022asf}.

The decays of $H_{\mathrm{V}}$ were studied using the sum rule method in
numerous publications. Thus, in Ref.\ \cite{Chen:2010ic} the authors used
the three-point correlation function for the vertex $H_{\mathrm{V}}\rho \pi $
and considered it in the limits $q^{2}\rightarrow 0$ ($q_{\mu }$ is a
momentum of the pion) and $m_{\pi }\rightarrow 0$. A pion-hybrid coupling
was extracted from analysis of a singular term $\sim 1/q^{2}$ in the
correlator and used later to estimate the partial width of the decay $H_{%
\mathrm{V}}\rightarrow $ $\rho \pi $. The width of this decay was found
around $180~\mathrm{MeV}$. The same approach allowed ones to evaluate widths
of the decays $H_{\mathrm{V}}\rightarrow $.$f_{1}\pi $ and $b_{1}\pi $ with
results $24~\mathrm{MeV}$ and $3~\mathrm{MeV}$, respectively. The widths of
the decays to $\eta \pi $ and $\eta ^{\prime }\pi $ mesons were found equal $%
<$ $1~\mathrm{MeV}$.

In the framework of the QCD light-cone sum rule approach the authors\ of
Ref.\ \cite{Huang:2010dc} in the case $m_{\mathrm{H}}=1.6~\mathrm{GeV}$
predicted: $\Gamma \lbrack \rho \pi ]=73\sim 120~\mathrm{MeV}$, $\Gamma %
\left[ f_{1}\pi \right] =69\sim 122~\mathrm{MeV}$, and $\Gamma \left[
b_{1}\pi \right] =0.14~\mathrm{MeV}.$ In the case of $m_{\mathrm{H}}=2.0~%
\mathrm{GeV}$ the results obtained there read: $\Gamma \lbrack \rho \pi
]=216\sim 370~\mathrm{MeV}$, $\Gamma \left[ f_{1}\pi \right] =109\sim 195~%
\mathrm{MeV}$, and $\Gamma \left[ b_{1}\pi \right] =3.7~\mathrm{MeV}$,
respectively. Later these estimates were modified and for $m_{\mathrm{H}%
}=2.0~\mathrm{GeV}$ became $\Gamma \lbrack \rho \pi ]=0.04~\mathrm{MeV}$ and
$\Gamma \left[ b_{1}\pi \right] =52-152~\mathrm{MeV}$ \cite{Huang:2016upt}.

The decay modes of the hybrid meson $H_{\mathrm{V}}$ were also considered
using alternative methods. In the lattice simulations \cite{McNeile:2006bz}
the mass of the hybrid meson ${1^{-+}}$ was found equal to $(2.2\pm 0.2)~%
\mathrm{GeV}$. In this work the partial widths of the processes $H_{\mathrm{V%
}}\rightarrow b_{1}\pi $ and $H_{\mathrm{V}}\rightarrow f_{1}\pi $ \ were
estimated $\Gamma \left[ b_{1}\pi \right] =(400\pm 120)~\mathrm{MeV}$ and $%
\Gamma \left[ f_{1}\pi \right] =(90\pm 60)~\mathrm{MeV}$. In other words,
this model implies fulfilment of the approximate ratio $\Gamma \left[
b_{1}\pi \right] /\Gamma \left[ f_{1}\pi \right] \rightarrow $\ $4/1$. \ The
flux tube model gave, at the same time, $\Gamma \left[ b_{1}\pi \right]
\approx 80~\mathrm{MeV}$ and $\Gamma \left[ f_{1}\pi \right] \approx 25~%
\mathrm{MeV}$ \cite{Burns:2006wz}, respectively.

Comparing the mass $2.30_{-0.17}^{+0.18}~\mathrm{GeV}$ of the hybrid state $%
H_{\mathrm{V}}$ from Ref.\ \cite{Barsbay:2024vjt} with available
experimental data, we see that it is considerably heavier than the structure
$\pi _{1}(1600)$. Its mass is close to the mass of the resonance $\pi
_{1}(2015)$. At the same time, the width of $H_{\mathrm{V}}$ is smaller than
the width $\Gamma =(230\pm 32\pm 73)~\mathrm{MeV}$ of the hybrid candidate $%
\pi _{1}(2015)$ presented in Ref.\ \cite{E852:2004rfa}. But $\Gamma \left[
H_{\mathrm{V}}\right] $ may be refined to reduce this gap by considering its
other decay channels. Therefore the hybrid $H_{\mathrm{V}}$ with some
caution may be interpreted as the meson $\pi _{1}(2015)$. It is interesting
that in Ref.\ \cite{Dudek:2010wm} $\pi _{1}(2015)$ was considered as
radially excited state of the meson $\pi _{1}(1600)$.  In Ref.  \cite{Eshraim:2020ucw},  the main decay modes  and the decay ratios  of light hybrid mesons including $\pi _{1}(1600)$ are also  investigated  in  the framework of the extended linear sigma model.

Our detailed theoretical analyses in the present study show  that the main decay channel of $H_{%
\mathrm{V}}$ is the process $H_{\mathrm{V}}\rightarrow \rho \pi $. This is
in accord with results of the sum rule analyses carried out in  Refs.\ \cite%
{Chen:2010ic,Huang:2010dc} though numerical predictions differ from our
findings. In the flux tube model  \cite{Burns:2006wz} and  expectation of extended linear sigma model \cite{Eshraim:2020ucw}, the main channel is decay to $b_{1}\pi $
mesons. Even from this non exhaustive information about the dominant decay
channels and full widths of the vector hybrid state $H_{\mathrm{V}}$
obtained by means of different methods, it is evident that our knowledge
about $H_{\mathrm{V}}$ is far from being complete. Existing discrepancies
between different approaches, or controversial results extracted in the
context of the same method demonstrate that a lot of work should be done to
reach a reliable conclusion about nature of this and similar exotic states.

Further experimental studies of the processes where the states $\pi
_{1}(1600)$ and $\pi _{1}(2015)$ were discovered are necessary. Such
investigations are required to clarify a situation with the mesons $\pi
_{1}(1400)$ and $\pi _{1}(1600)$ and may lead to observation of new hybrid
candidates.


\section*{ACKNOWLEDGMENTS}

G.~Daylan~Esmer, H.~Sundu and S.~T\"{u}rkmen are thankful to Scientific and
Technological Research Council of T\"{u}rkiye (TUBITAK) for the financial
support provided under the Grant No. 123F197. The authors thank S.~S.~Agaev
for useful discussions and comments.

\end{document}